\documentclass[aps,showpacs,showkeys,preprintnumbers,twocolumn,
groupedaddress,nofootinbib,eqsecnum,prd]{revtex4-1}

\usepackage{amsmath,amssymb}
\usepackage{graphicx}
\usepackage{xspace}

\usepackage[usenames,dvipsnames]{color}
\usepackage{url}
\usepackage[pdftex,%
  colorlinks=true,
  urlcolor=blue,               
  anchorcolor=blue,
  filecolor=green,             
  linkcolor=red,               
  menucolor=blue,
  citecolor=Sepia,
  breaklinks=true,
  bookmarks=true]{hyperref}

\def\gsim{\mathrel{\rlap{\raise 2.5pt \hbox{$>$}}\lower 2.5pt\hbox{$\sim$}}}
\def\lsim{\mathrel{\rlap{\raise 2.5pt \hbox{$<$}}\lower 2.5pt\hbox{$\sim$}}}

\def\stilde{\widetilde}

\newcommand{\Bmumu}{\ensuremath{\mathrm{BR}(B_s \to \mu^+ \mu^-)}\xspace}
\newcommand{\Bsg}{\ensuremath{\mathrm{BR}(B \to X_s \gamma)}\xspace}
\newcommand{\Btn}{\ensuremath{\mathrm{BR}(B^+ \to\tau^+\nu_\tau)}\xspace}
\newcommand{\Cha}[1]{\ensuremath{\stilde \chi_{#1}^\pm}\xspace}
\newcommand{\Chap}[1]{\ensuremath{\stilde \chi_{#1}^+}\xspace}
\newcommand{\Cham}[1]{\ensuremath{\stilde \chi_{#1}^-}\xspace}
\newcommand{\Neu}[1]{\ensuremath{\stilde \chi_{#1}^0}\xspace}
\newcommand{\bsg}{\ensuremath{b \to s\gamma}\xspace}
\newcommand{\gmu}{\ensuremath{(g-2)_\mu}\xspace}
\newcommand{\mgl}{\ensuremath{m_{\tilde g}}\xspace}
\newcommand{\mhalf}{\ensuremath{M_{1/2}}\xspace}
\newcommand{\mst}{\ensuremath{m_{\tilde t_1}}\xspace}

\newcommand{\refeq}[1]{Eq.~\eqref{#1}}
\newcommand{\stp}{\ensuremath{\tilde t_1}\xspace}
\newcommand{\tb}{\ensuremath{\tan\beta}\xspace}

\begin{document}

\preprint{HIP-2011-17/TH}

\title{Stop as a next-to-lightest supersymmetric particle in constrained MSSM}

\author{Katri Huitu}
\email{katri.huitu@helsinki.fi}
\author{Lasse Leinonen}
\email{lasse.leinonen@helsinki.fi}
\affiliation{Department of Physics, and Helsinki Institute of Physics,
  FIN-00014 University of Helsinki, Finland}
\author{Jari Laamanen}
\email{j.laamanen@science.ru.nl}
\affiliation{Theoretical High Energy Physics,
  IMAPP, Faculty of Science,  Radboud University Nijmegen,
  Mailbox 79,
  P.O. Box 9010,
  NL-6500 GL Nijmegen, The Netherlands}


\begin{abstract}

  So far the squarks have not been detected at the LHC indicating that
  they are heavier than a few hundred GeVs, if they exist.  The
  lighter stop can be considerably lighter than the other squarks.  We
  study the possibility that a supersymmetric partner of the top
  quark, stop, is the next-to-lightest supersymmetric particle in the
  constrained supersymmetric standard model.  Various constraints, on
  top of the mass limits, are taken into an account, and the allowed
  parameter space for this scenario is determined.  Observing stop
  which is the next-to-lightest supersymmetric particle at the LHC may
  be difficult.

\end{abstract}


\maketitle


\section{Introduction} 
The next-to-lightest supersymmetric particle (NLSP) has a crucial role
in the attempts of detecting supersymmetry (SUSY) since the lightest
supersymmetric particle will escape the detectors unnoticed.  The
Large Hadron Collider (LHC) accelerates and collides mainly protons,
and thus the collision processes are overwhelmed by the strong
interactions.  The superpartner of the top quark, stop, can be the
lightest colored superpartner, mainly due to the splitting of the two
stop scalar states amplified by the top Yukawa coupling.

In principle, the masses of the superpartners are free parameters in
the minimal supersymmetric extension of the standard model (MSSM). The
large number of the parameters parametrize the supersymmetry breaking,
which is expected to be spontaneous in a more complete theory.  The
explicit SUSY breaking is introduced softly so that no quartic
divergences re-appear. This requires inclusion of all the possible
breaking terms, which are gauge invariant, into the Lagrangian.  In
models of a particular SUSY breaking mechanism, the number of
parameters may be substantially smaller.  Perhaps the most studied
model is the constrained minimal supersymmetric standard model (CMSSM)
\cite{Chamseddine:1982jx,Kane:1993td}, which we also consider here.
In CMSSM, supersymmetry is supposed to be broken spontaneously at the
hidden sector by the SUSY breaking fields, which do not have gauge
interactions with the SUSY fields that are, in principle, observable
to us.

The breaking fields affect us through the gravitational strength
interactions, which generate an effective Lagrangian with SUSY
breaking terms. The effective Lagrangian does not have to be
renormalizable anymore, so it contains terms with couplings to the
hidden sector fields suppressed by powers of inverse Planck mass.
After the hidden sector fields generate vacuum expectation values,
(renormalizable) soft SUSY breaking terms arise. This implies, under
certain minimality assumptions, that the SUSY scalar terms have common
couplings: the common scalar mass parameter $m_0^2$, a common bilinear
parameter $B_0$, as well as a common trilinear parameter $A_0$, at the
energy scale where the Lagrangian is established. This scale is
thought to be the grand unification (GUT) scale.

Since the observations are made at the low energies, the
renormalization group equations (RGE) must be used to calculate the
running parameters, like the sparticle masses, at the low-energy
scale.
Even though the scalars have equal mass parameters at the GUT scale,
this is not true at the electroweak (EW) scale anymore. Each scalar
RGE has terms proportional to the corresponding fermionic partner
Yukawa coupling. The effect of these terms is to decrease the mass
parameter in question. Because of the large size of the top Yukawa
coupling, the corresponding mass parameters entering to the stop mass
matrix tend to decrease most, leading to the situation where the stops
are the lightest squarks. The terms with gauge couplings have an
opposite effect, as they appear with an opposite sign.  Since the
slepton mass parameter RGEs lack terms with the strong coupling
constant, they do not increase as much as the squark mass
parameters. Therefore, both the lighter stop and the lightest slepton,
stau, are natural candidates for being the lightest supersymmetric
scalar.

Supersymmetry by itself does not prevent the introduction of baryon
and lepton number violating terms, which are not present in the
standard model (SM) renormalizable Lagrangian. Such terms have a
potential to lead to a very fast proton decay. Their presence is
prevented by so-called $R$-parity \cite{Salam:1974xa, *Fayet:1974pd,
  *Farrar:1978xj, *Dimopoulos:1981dw, *Farrar:1982te}: SM particles
are positively, and their superpartners, sparticles, negatively
charged under this parity.  The remarkable consequence is that the
lightest supersymmetric particle (LSP) must be absolutely stable.
Therefore, a large number of LSPs may still be around from the early
universe, which could explain the observed dark matter relic density
(RD).  In particle colliders, sparticles can be produced only in even
numbers, since the initial state contains only ordinary particles. The
LSP is thought to be uncharged (in both electric and color charges)
which means that it may be only weakly interacting or that it only has
gravitational interactions. Therefore, it is expected that it escapes
the detectors unnoticed; only a missing energy component transverse to
the colliding beam, $E^{\mathrm{miss}}_T$ (MET), is noticed.

The obvious candidate for the LSP in the MSSM spectrum is the lightest
neutralino, \Neu 1, which is the mixture of the neutral higgsinos,
bino and wino, {\it i.e.} the superpartners of Higgs, $B$ and
$W$-bosons, respectively.  Other candidates are the superpartner of
graviton, gravitino, and superpartner of a neutrino, sneutrino.
In gauge mediated supersymmetry breaking (GMSB) models
\cite{Dine:1993yw,*Dine:1994vc,*Dine:1995ag}, the LSP is always the
gravitino, which is thus a natural dark matter candidate in these
models.  In GMSB, the leading contribution to the squark masses is
proportional to the strong coupling constant, being larger than the
leading contribution to the slepton masses.  Thus, it is not expected
that a strongly interacting particle is the NLSP in a minimal GMSB
model.
The searches at LEP collider exclude the possibility of light
left-handed sneutrinos as the LSP, and masses beyond LEP's reach are
ruled out by direct detection dark matter searches
\cite{Caldwell:1988su,*Caldwell:1990tk,*Reusser:1991ri,*Mori:1993tj}.
Unlike the partner of the left-handed neutrino, the partner of the
right-handed neutrino is a viable dark matter candidate
\cite{Asaka:2005cn,*Asaka:2006fs}.  In such an extended model, also
signatures of stop NLSP have been studied at the colliders
\cite{deGouvea:2006wd,*Choudhury:2008gb}.

In this paper, we study, within CMSSM, the possibility that the
lighter stop scalar state is the NLSP. This, in particular, states
that the lightest stop \stp is lighter than the lighter chargino (\Cha
1), the mixture of charged higgsinos and winos\footnote{In the
  anomaly mediated supersymmetry breaking (AMSB) models
  \cite{Randall:1998uk,*Giudice:1998xp,*Gherghetta:1999sw,*Pomarol:1999ie}
  chargino is the NLSP.}.
This also implies that the LSP is a neutralino\footnote{The gravitino
  mass is not fixed in CMSSM, but the gravitino LSP with stop NLSP is
  not allowed \cite{DiazCruz:2007fc}. In the so-called NUHM models
  \cite{Baer:2005bu,DiazCruz:2007fc} stop can be the NLSP while
  gravitino is the LSP.}.  This kind of a scenario requires large
splitting between the two stop scalars. One consequence is that a wide
gap opens in the sparticle mass spectrum between the scalars. \stp is
the lonely scalar and close to the mass of \Neu 1 as a result of the
relic density constraint. Therefore, stop is supposed to be the
sparticle that is produced at the LHC in abundance.
Stop NLSP in CMSSM with nonzero trilinear term has been studied, for
example, in
Refs.~\cite{Chattopadhyay:2007di,Gladyshev:2007ec,Das:2001kd}.
Implications at colliders have been studied, e.g., in
Refs.~\cite{Bhattacharyya:2008tw,Bornhauser:2010mw,Johansen:2010ac,Hiller:2009ii,Kraml:2005kb,Das:2001kd,Olive:1994qq,Demina:1999ty}.
In this work, we scan over all parameters and thus will
comprehensively determine the possible regions of stop NLSP.  We also
map all the regions where dark matter constraints are met.
Stop NLSP may also arise in some compressed SUSY models
\cite{Martin:2007gf,*Martin:2008aw,Baer:2007uz} or in the context of
mirage mediation scenario
\cite{Choi:2005uz,*LoaizaBrito:2005fa,*Choi:2006im,Huitu:2010me}.
Light stop scenario is also favored in the view of the $b-\tau$ Yukawa
unification \cite{Gogoladze:2011be}.

In section \ref{sec:stop-as-an}, we discuss the conditions for stop
being the NLSP and the decay modes of the stop NLSP.  In section
\ref{sec:allconstrainst}, we list the constraints used in this paper.
In section \ref{sec:there-are-points}, we scan over relevant
parameters to map the possible stop NLSP regions.  In section
\ref{sec:at-lhc}, we discuss the possibilities for detection of stop
NLSP at the LHC, and we conclude in section \ref{summary}.

\section{Stop as an NLSP}
\label{sec:stop-as-an}
Stop can be the NLSP when the mixing term of the left- and
right-labeled scalar states $M^2_{LR}=v(a_t\sin\beta-\mu y_t
\cos\beta)$ in the stop mass matrix is large enough. The largest
mixing occurs when the supersymmetric Higgs mass parameter $\mu$ and
the trilinear SUSY breaking parameter $a_t$ have opposite signs.  Here,
$y_t$ is the top Yukawa coupling, $\beta$ is defined in the relation
$\tb = v_2/v_1$, $v_i$ being the vacuum expectation values of the two
Higgs doublets obeying the relation $v=\sqrt{v_1^2+v_2^2}$.
We assume here that the trilinear soft parameters are proportional to
the Yukawa couplings, $a_i = y_i A_i$, so the above mixing term can be
written as $M^2_{LR}=m_t (A_t -\mu \cot\beta)$.

In addition to the mixing, the renormalization group running plays an
important role in the determination of the NLSP. In the scalar RGEs,
the terms proportional to the Yukawa couplings (and the scalar masses)
decrease the soft mass parameters, while the terms proportional to the
gauge couplings (and gaugino masses) have an opposite effect. Because of
the largeness of the strong coupling constant, the squark mass
parameters tend to increase more than the slepton mass parameters,
even though the third-generation squarks have large Yukawa
couplings. Therefore, a small gaugino mass parameter as compared to the
scalar mass parameter is preferred in order to suppress the strong
coupling term in the squark RGE, which then may lead to a stop NLSP
instead of stau NLSP.

The mass of the NLSP affects the relic density through
co-annihilations. (The neutralino-stop co-annihilations were studied
in Ref.~\cite{Ellis:2001nx}.) Unless the annihilating LSPs are close to
a resonance or are light, their annihilation usually cannot result in
dark matter abundance within the observed limits.  Rather, the relic
density is usually too large.  Co-annihilations with other particles,
however, dilute the relic density, and with certain parameters,
annihilations can even be too effective so that there is hardly any
dark matter left after the annihilations cease. In the case of large
(negative) values of $A_t$, $\cal{O}$(TeV), and stop NLSP, the mass
difference between stop and the neutralino LSP should not be less than
20 GeV or more than around 50 GeV in order to obtain the desired relic
density, as we will show later (in the case of positive $\mu$).

\begin{figure}
  \centering
  \includegraphics{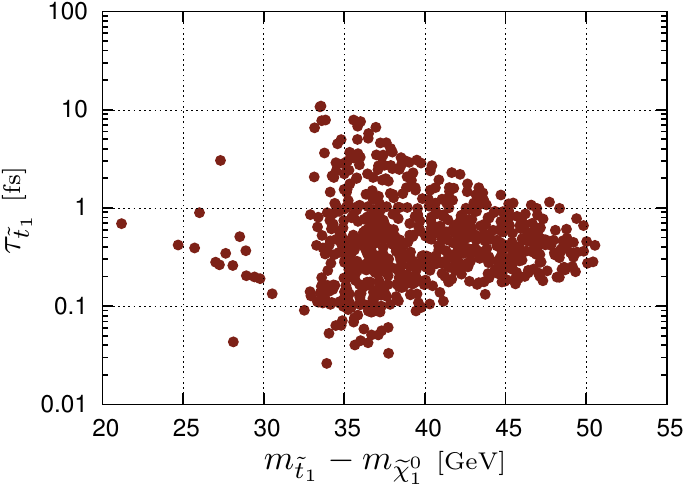}
  \caption{Stop lifetime [fs] as a function of the NLSP-LSP mass
    difference. The points are the same RD=WMAP points as used later
    in Fig.~\ref{fig:mst-lsp_stNLSPoksLEQ}.}
  \label{fig:lifetime}
\end{figure}

When stop is the NLSP and $m_{\tilde{t}_1} - m_{\tilde{\chi}^0_1} <
m_W$, the only possible stop decays are
\begin{itemize}
\item $\tilde{t}_1 \rightarrow u \tilde{\chi}^0_1$
\item $\tilde{t}_1 \rightarrow c \tilde{\chi}^0_1$
\item $\tilde{t}_1 \rightarrow b f\bar{f}^{\prime} \tilde{\chi}^0_1$.
\end{itemize}
Because of the required mass difference, the lifetime of stop is short, of
the order of $10^{-15} s$ (Fig.~\ref{fig:lifetime}).
The loop decay $\tilde{t}_1 \rightarrow c \tilde{\chi}^0_1$ is
enhanced by a large logarithm
$\mathrm{ln}\,(\Lambda^2_{\mathrm{GUT}}/m^2_W)$, and is likely to be
dominant over the four-body decay, while smaller
Cabibbo-Kobayashi-Maskawa (CKM) matrix elements suppress the other
loop decay \cite{Hikasa:1987db}.
There are also scenarios where the four-body decay rate is at the same
level or larger than the loop decay, for example when squark masses
are not unified at a large mass scale, resulting only in a small
logarithmic enhancement \cite{Boehm:1999tr}.
In section~\ref{sec:at-lhc}, we will discuss the decay modes of stop
when combined with the constraint requirements.

\section{Constraints}
\label{sec:allconstrainst}
In order to test the validity of a model, a number of constraints must
be considered.

\subsubsection{The Higgs boson mass and other direct searches}
\label{sec:hmass}
The nonobservation of Higgs bosons and supersymmetric partners put
stringent constraints on the parameters of SUSY models.  The 95\%
C.L. exclusion limit for the SM Higgs boson mass of $m_H > 114.4$ GeV
\cite{Nakamura:2010zzi} is applicable also to the lightest CP-even
Higgs boson of the MSSM if it is SM-like. This implies that the Higgs
coupling to the $Z$ boson stays near to the SM value. If the
$hZZ$-coupling is reduced, as may happen, e.g., at the large \tb, a
general SUSY Higgs mass bound, $m_h > 92.8$ GeV, must be applied
\cite{Nakamura:2010zzi}.

The light Higgs boson mass in SUSY models is generally sensitive to
the higher-order corrections. Since there is a theoretical uncertainty
of some 3 GeV from the next-to-next-to-leading order (NNLO) and higher
corrections \cite{Degrassi:2002fi,*Heinemeyer:2004ms}, a limit of $m_h
> 111$ is used in our calculations instead of the kinematic bound of
114.4 GeV.  The SM-like $Z$-coupling is checked for this bound: A
value $\sin^2(\beta-\alpha_{eff}) > 0.9$ is required for the $m_h >
111$ GeV limit in our work, otherwise, a limit $m_h > 91$ GeV is used,
which also leaves some room for theoretical uncertainties.  Here,
$\alpha_{eff}$ is the effective mixing angle between neutral Higgs
bosons.

The lightest Higgs boson mass is limited by $m_Z$ at the tree level
\cite{Inoue:1982ej} even for more than two Higgs doublets
\cite{Flores:1982pr}. The Higgs mass depends strongly on the \tb, and
a large \tb is favored. The radiative corrections are crucial for
bringing the light Higgs mass up to the acceptable level. The largest
contributions typically come from the top and stop loops. In the
decoupling regime ($m_A \gg m_Z$ with $\tb \gg 1$), the leading 1-loop
radiative correction to $m_h^2$ can be written as $m_h^2=m_Z^2+
\epsilon_t+\epsilon_b$ \cite{Allanach:2004rh}, where
\begin{align}
  \epsilon_t &= \frac{3 m_t^4}{2 \pi^2 v^2} \left( \ln
    \frac{M_S^2}{m_t^2} + \frac{X_t^2}{2\,M_S^2} - \frac{X_t^4}{12
      M_S^4}
  \right), \label{etop}\\
  \epsilon_b &= -\frac{3 m_b^4}{2 \pi^2 v^2} \frac{X_b^4}{12 M_S^4}.
  \label{epsmix}
\end{align}
Here, $M_S$ is a common soft SUSY--breaking mass term for the
third--generation squarks and $X_{t,b}$ are the mixing terms involving
$\mu$, $\tb$, and the stop/sbottom trilinear couplings,
\begin{align}
  X_t&= A_t-\mu\cot\beta,\label{eq:5}\\
  X_b&= A_b - \mu\tb.\label{eq:6}
\end{align}
Because of the quartic dependence on the top mass, the $\epsilon_t$
term is the leading one. In the maximal mixing scenario
\cite{Carena:1999xa}, the second and third terms of \refeq{etop}
cancel each other, leading to the mixing condition
\begin{equation}
  \label{eq:1}
  X_t = \sqrt{6} M_S.
\end{equation}
In the case of stop NLSP, the Higgs mass constraint seems to require
nearly maximal mixing, as can be seen later.

Also, the other collider bounds from LEP, like chargino mass limits,
and the LHC squark and gluino limits (CMS \cite{Khachatryan:2011tk},
ATLAS \cite{Collaboration:2011qk}) are important constraints. The LEP
limits have been taken into account, as implemented in the program
micrOmegas (v.2.4.R)~\cite{Belanger:2006is}.  The new LHC limits were
also considered, where applicable.

\subsubsection{The $B \to \tau\nu$ decay}
\label{sec:b-to-taunu}

The final states of the decays of the type $B \to l\nu$ consist purely
on leptonic states, and the hadronic uncertainties are present only in
the $B$ meson decay constant $f_B$. Only the $\tau$ channel has been
observed so far. The SM expected value for the branching ratio, when
using the value of $|V_{ub}|$ given by the UTFit collaboration
\cite{Bona:2009cj}, is predicted to be \cite{Bhattacherjee:2010ju}
\begin{equation}
  \label{eq:smPred-btn}
  \mathrm{BR}(B^+ \to\tau^+\nu_\tau)_\mathrm{SM}
  = (0.80\pm 0.15)\times 10^{-4} .
\end{equation}
Recently, the experimental measurements from the $B$-factories have
improved significantly, the most recent world average measurement
being \cite{JampensTalk}
\begin{equation}
  \label{avbtn}
  \mathrm{BR}(B^+ \to \tau^+ \nu_\tau)_\mathrm{exp}
  = (1.68 \pm 0.31) \times 10^{-4}.
\end{equation}
The new physics (NP) contribution to the branching ratio can be quantified
by defining a ratio \cite{Isidori:2006pk}
\begin{equation}
  R^{\mathrm{NP}}_{\tau\nu_\tau}\equiv\frac{\mathrm{BR}(B^+\to\tau^+
    \nu_\tau) _{\mathrm{\,SM+NP}} }
  {\mathrm{BR}(B^+\to\tau^+\nu_\tau)_{\mathrm{SM}}},
\end{equation}
where the subscript SM+NP denotes the branching ratio of the NP
scenario, including the SM contribution. The 95$\%$ C.L. allowed range
for $R^{\mathrm{NP}}_{\tau\nu_\tau}$ is then \cite{Bhattacherjee:2010ju}
\begin{equation}
  0.99 < R^{\rm{NP}}_{\tau\nu_\tau} < 3.19.
  \label{RatioExp}
\end{equation}
With higher-order corrections, the formula for the ratio is
\cite{Akeroyd:2003zr,Isidori:2006pk}
\begin{equation}
  R^{\mathrm{NP}}_{\tau\nu_\tau} =
  \left(1-\frac{\tan^2\beta}{1+\tilde\epsilon_0\,\tan\beta}\,\frac{m^2_B}
    {M^2_{H^\pm}}\right)^2,
  \label{RatioEps}
\end{equation}
where $\tilde\epsilon_0$ contains all the higher-order corrections and
$m_B$ is the $B$-meson mass. The branching ratio depends strongly on
the charged Higgs mass and \tb, and typically the constraint
\eqref{RatioExp} prefers small values of \tb in order not to decrease
the ratio too much below the lower limit (unless $M_{H^\pm}$ is very
light). A large charged Higgs mass decreases the new physics
contributions in general.

\subsubsection{The $(g-2)$ of muon}
\label{sec:g-2}
The anomalous magnetic moment of muon, $a_{\mu}=(g-2)/2$ has been
measured quite precisely a decade ago.  The measured value
\cite{Bennett:2006fi} for $a_{\mu}$  is
\begin{equation}
  \label{eq:amuexp}
  a_{\mu}^{\text{exp}}= (11659208.0 \pm 6.3)\times 10^{-10}.
\end{equation}
The recent SM prediction for this is \cite{Miller:2007kk}
\begin{equation}
  a_{\mu}^{\text{SM}}= (11659178.5 \pm 6.1)\times 10^{-10}.
\end{equation}
leading to a discrepancy between the SM and experiment\footnote{In a
  very recent paper \cite{Bodenstein:2011qy} it was argued that the
  experimental and SM value actually agree.},
\begin{align}
  \label{eq:2}
  {\Delta a}_{\mu}
  &=(29.5 \pm 8.8) \times 10^{-10}.
\end{align}
This is a $3.4~\sigma$ deviation.  The SM prediction is largely
dominated by leptonic QED processes, though careful calculation of
hadronic and electroweak contributions is also necessary due to high
precision of the experimental measurement.  The hadronic processes,
vacuum polarization above all, have more than an order of magnitude
larger contribution to the magnetic moment compared to the electroweak
processes.

Purely supersymmetric contribution of MSSM to $a_{\mu}$ is
proportional to $\tan \beta \,\mathrm{sign}(\mu)/M^2_{\mathrm{SUSY}}$,
and for large enough $\tan \beta$ and not too heavy supersymmetric
particles, it can be larger than the electroweak contribution. For
positive values of $\mu$, the MSSM can provide the solution to the
discrepancy.  For negative values of $\mu$, the new physics
contributions drive the gap even wider. Because of this, a negative sign
for $\mu$ in the MSSM is usually considered to be disfavored
\cite{Stockinger:2006zn}.

As pointed out in \cite{Bhattacherjee:2010ju}, this constraint is
complementary to the $B \to \tau\nu$ decay constraint, and taken
together, they rule out large areas of the parameter space.

The theoretical calculation of $a_\mu$ is known to be difficult because of
the hadronic contributions and nonperturbative effects involved, see
{\it e.g.}  \cite{Knecht:2003kc,Prades:2009qp,Hagiwara:2011af} and
references therein.  Although there has been impressive improvements
in the calculation, frequently the $a_\mu$ constraint is not used in
determining the excluded parameter space.  We will take a similar
attitude here, but comment on $a_\mu$ when appropriate.
When referring to this constraint, the following 95\% C.L. limits,
which include theoretical uncertainties, are used
\cite{Eriksson:2008cx}:
\begin{equation}
  \label{eq:3}
  1.15 \times 10^{-9} 
  < \Delta a_\mu^{SM+NP}< 
  4.75 \times 10^{-9}.
\end{equation}
Our acceptable parameter points have $\Delta a_\mu^{SM+NP}$ below this
range.

\subsubsection{The $b \to s\gamma$ branching ratio}
\label{sec:b-to-sgamma}
The present experimental value by the Heavy Flavor Averaging Group
(HFAG) is \cite{Asner:2010qj}
\begin{eqnarray}
 BR(B \to X_s \gamma) &=& (355 \pm 24 \pm
9) \times 10^{-6}.\nonumber
\end{eqnarray}
In our constraint, the 
theoretical uncertainties are included as well: in the SM at the NNLO QCD
level the uncertainty can be estimated to be $23 \times 10^{-6}$
\cite{Misiak:2006zs}, in the MSSM, the theoretical uncertainty is estimated
to be additional 5\% (we take $15 \times 10^{-6}$)
\cite{Ellis:2007fu}. Combining all these gives (at $2~\sigma$)
\begin{equation}
  \label{eq:4}
  BR(B \to X_s \gamma) = (355 \pm 142) \times 10^{-6}.
\end{equation}
The \bsg constraint is sensitive to the sign of $\mu$
\cite{Nath:1994tn}, preferring the positive value.

\subsubsection{Other constraints}
\label{sec:b_s-to-mumu}
The $B_s \to \mu^+ \mu^-$ branching ratio can also be used as a
constraint for new physics. The most recent experimental upper limit
by the CDF collaboration is \cite{d.:_searc_rare_decay_bs}
\begin{equation}
  \label{eq:expBmumu}
  BR(B_s \to \mu^+ \mu^-) < 4.3 \times 10^{-8}\ (95\% \text{ C.L.})
\end{equation}
Including the theoretical uncertainty from $f_{Bs} = 238.8 \pm 9.5$
MeV \cite{Laiho:2009eu}, we find a conservative upper limit of
\begin{equation}
  \label{eq:Bmumulim}
  BR(B_s \to \mu^+ \mu^-) < 5.0 \times 10^{-8}.
\end{equation}
In practice, this constraint is usually not the limiting factor, and
in our calculations it is never the main constraint in otherwise good
points. Therefore, we do not comment on this constraint further.

\subsubsection{Relic density}
\label{sec:relic-density}
Stable, (color and charge) neutral LSP provides a convenient candidate
to explain the dark matter abundance observed by the Wilkinson
Microwave Anisotropy Probe (WMAP) satellite mission.  With the
combined data from 7-year WMAP results, BAO (Baryon Acoustic
Oscillations) in the distribution of galaxies, and observation of
Hubble constant, the density of cold dark matter in the universe is
determined to be \cite{Komatsu:2010fb} $\Omega_c h^2 = 0.1126 \pm
0.0036$.  If 10 \% theoretical uncertainty is added
\cite{Baro:2007em}, we find the preferred WMAP range of
\begin{equation}
  \label{wmaplimits}
  0.0941
  < \Omega_c h^2 <
  0.1311
\end{equation}
at  $2~\sigma$ level.
CMSSM can provide a dark matter density that is within the WMAP
limits, at least in some parts of the parameter space.  However, dark
matter may be of nonsupersymmetric origin, so here we have used only
the upper bound as a real constraint, unless otherwise indicated.

\section{Stop NLSP parameter space}
\label{sec:there-are-points}

\subsection{Method}
\label{sec:method}

The CMSSM parameter space is in practice four dimensional
($m_0,\mhalf,A_0,\tb$), with the sign ambiguity of the $\mu$-parameter
doubling it. The actual value of the $\mu$-parameter is calculated
from the radiative electroweak symmetry breaking (rEWSB) constraint
condition
\cite{Inoue:1982ej,Ibanez:1982fr,*AlvarezGaume:1983gj,*Ellis:1982wr},
which leaves the sign of $\mu$ as a free parameter. \mhalf is the common
gaugino mass, defined at the GUT scale by the boundary condition
$\mhalf = M_1 = M_2 = M_3$, where $M_{1,2,3}$ are the $U(1)$, $SU(2)$ and
$SU(3)$ gaugino masses, respectively.
Limiting oneself to 2-dimensional parameter scans, on the
one hand, reduces the effect of the other parameters to discrete examples,
and on the other hand, creates easily an illusion of very restricted
parameter space volume. Therefore, it is useful to actually explore
the full parameter space.
In order to find the parameter points, where stop is the NLSP
and the experimental constraints are not violated, we scanned the
4-dimensional parameter volume of the three GUT scale parameters and
$\tb$ and calculated the low energy particle spectrum and the
constraints for each point. The sign of $\mu$ was chosen to be
positive, partly guided by the preferences for some of the constraints
of Sec.~\ref{sec:allconstrainst}. It should be noted that we
restricted ourselves to the selection of real-valued parameters, which
implies that no additional CP-violation is introduced by the soft SUSY
breaking terms.

The particle spectrum was calculated using SOFTSUSY
(v.3.1.7)~\cite{Allanach:2001kg}, and the relic density and
constraints using micrOmegas (v.2.4.R)~\cite{Belanger:2006is}. Top
pole mass of $m_t = 173.3$ GeV was used throughout this study.

\subsection{Stop NLSP regions}
\label{sec:stop-nlsp-regions}

\begin{figure}
  \centering
  \includegraphics{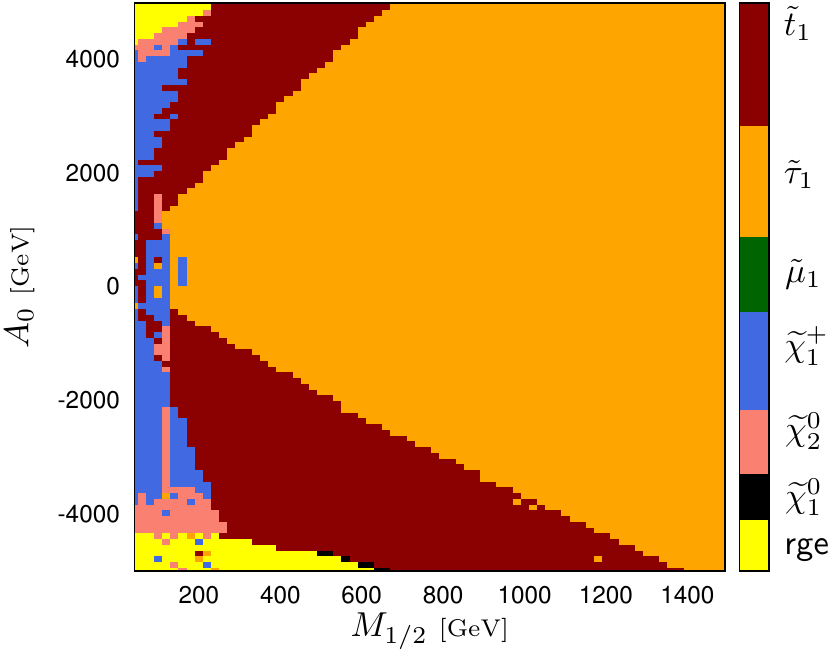}
  \caption{The potential stop NLSP parameter space in dark (brown) on
    $(\mhalf,A_0)$ plane. No constraints are applied yet.}
  \label{fig:stopmapm0a0Large}
\end{figure}

Figure~\ref{fig:stopmapm0a0Large} shows a 2-dimensional projection of
the scanned 4-dimensional parameter space,
where the ranges for parameters are (in [GeV], $\mu>0$):
\begin{eqnarray}
& 50 \leq \mhalf \leq 1490 | 20;&\  \nonumber\\
&50 \leq m_0 \leq 1990 | 20;&\  \nonumber\\
&-5000 \leq A_0 \leq 5000 | 100,& \nonumber\\
&\tb \in \{2...7|1; 10...50|10\}.&\nonumber
\end{eqnarray}
The number after $"|"$ denotes the stepping.
Each point shows the identity of one of the possibly many potential
NLSPs which results from varying the remaining two free parameters
(\tb and $m_0$ in this case). The representative NLSP identity is
chosen by keeping in mind that we would like to find all the points where
stop can be the NLSP. Therefore, if a stop NLSP is found, it is
chosen. Otherwise a stau, \Cha 1 or \Neu 2 NLSP is chosen, in that
order of preference. Thus, an area labeled with $\tilde\tau_1$ does
not allow stop NLSPs for the scanned parameters, but may
contain, e.g., \Neu 2 NLSP. The rge-denoted (yellow) area shows the
space where no combination of the parameters provides a good solution
to the spectrum calculation (tachyons, no rEWSB \ldots).
A large area of stop NLSP is found when the $A_0$-parameter is
nonzero. With our choice of positive $\mu$, the negative
$A_0$-parameter leads to larger mixing in stop mass matrix, see
\refeq{eq:5}. Therefore, for a symmetric range, a larger number of
potential stop NLSP points is found with negative $A_0$-parameters.
A large $\cot\beta$ is favored because of the enhancing effect
in the stop mixing on the one hand, and because of the suppressing
effect in the stau mixing on the other hand.  Since \tb is paired
with the $\mu$-parameter, which is determined from the rEWSB
condition, general upper or lower limits for \tb are very
involved. From our numerical calculations, however, we learn that
the stop NLSP exists up to $\tb=50$, but for $\tb=55$, stau is the
NLSP. For a large value of \mhalf, the upper value for \tb drops to
about 35 (stau being the NLSP otherwise).
To have a better view of the stop NLSP points, we look at the negative
$A_0$-parameters in more detail. We anticipate this also to alleviate
the lightest Higgs boson mass limit constraint, as discussed in
Sec.~\ref{sec:hmass}.  Some numerically unstable points may give
erroneously large $m_h$, hence, we set also an upper limit for $m_h$ to
be 140 GeV \cite{Kane:1992kq,Espinosa:1992hp,Allanach:2004rh}, which
cuts the unreliable points from the calculations.

\begin{figure}
  \centering
  \includegraphics[width=.5\textwidth]{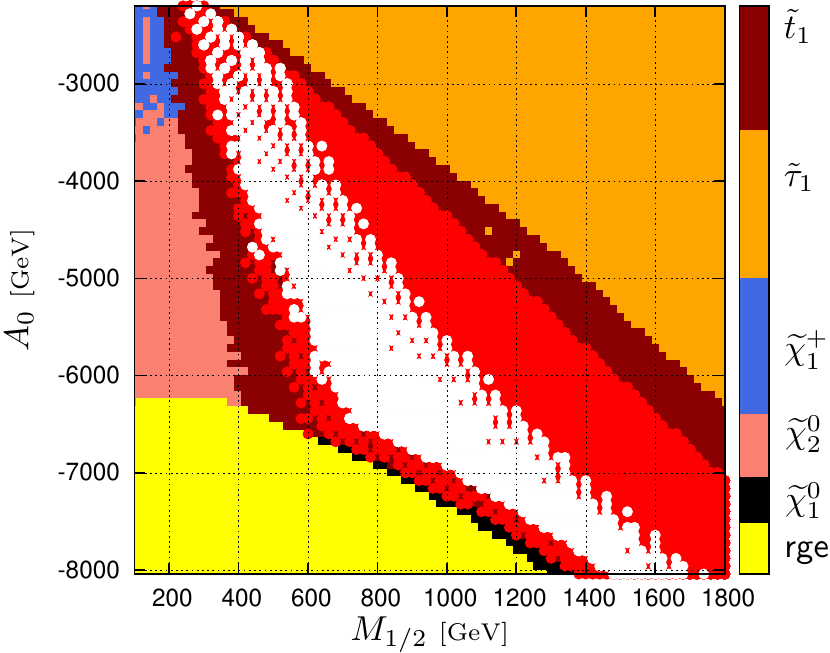}
  \caption{The potential stop NLSP area in $(\mhalf,A_0)$ plane.
    Zoomed from Fig.~\ref{fig:stopmapm0a0Large} to an interesting
    area. The red points are the stop NLSP points with relic density
    below the WMAP upper limit (which we denote from now on as RD
    $\leq$ WMAP), and the white subset of the points also obey the
    constraints of Sec.~\ref{sec:allconstrainst}. See the text.}
  \label{fig:stopOKsLTwmap}
\end{figure}

In Fig.~\ref{fig:stopOKsLTwmap}, the negative $A_0$-parameter space is
plotted with respect to \mhalf as a 2-dimensional projection of the
4-dimensional parameter space showing the NLSP map.  Parameters for
subsequent figures are (in [GeV], $\mu>0$):
\begin{eqnarray}
&100 \leq \mhalf \leq 1800 | 20;& \nonumber\\
&1000 \leq  m_0 \leq 2848 | 66;& \nonumber\\
& -8040 \leq A_0 \leq -2200 | 80;  & \nonumber\\
&\tb \in \{2.5...20|0.5; 25...55|5\}& \nonumber
\end{eqnarray}
A large space with stop NLSP is found. However, many of the potential
stop-NLSP points conflict with the constraints specified earlier.
The requirement of the neutralino relic density to fully explain the
observed WMAP cold dark matter density is quite a restrictive
constraint. This constraint may be weakened to the requirement that
neutralino dark matter relic density should not exceed the upper WMAP
limit for relic density.  This, however, means that we cannot explain
the dark matter puzzle with the SUSY model.
Figure~\ref{fig:stopOKsLTwmap} shows, on top of the NLSP map, also
scattered (red) points, where stop is the NLSP and the neutralino
relic density is below the WMAP upper limit. The white subset of these
points shows the parameters for which, in addition, the other
constraints of Sec.~\ref{sec:allconstrainst} are not violated
(however, excluding the muon magnetic moment, which is below the range
\eqref{eq:3}).  In other words, these points are not excluded by the
constraints discussed and exhibit the stop NLSP.
When examined one by one, the \Btn constraint limits the points only
near the rge-denoted area, which is true also for the \bsg
constraint. In contrast, the possible discrepancy in \gmu cannot be
explained with these models, since the calculated SUSY contribution
here is always below the range of \refeq{eq:3}. When applying all the
constraints (except \gmu) simultaneously, the white point area
appears.
It should be noted that there exists a subset of the above points,
which nicely match the WMAP constraint, thus explaining also the
amount of dark matter in the universe. Moreover, it matches the shown
white dot area, just being sparser due to the limited number of
scanned points. Therefore, it is possible to find points where stop is
the NLSP, neutralino relic density matches the WMAP observation, and
the collider constraints are fulfilled. This leaves still the possible
deviation of the observed \gmu unexplained.
Even though the \gmu value could be increased by considering the  
points with large \tb, this conflicts with the \Btn constraint. In  
addition, $|A_0|$ should not be larger than 2--3 TeV for the \gmu  
constraint to be fulfilled. \gmu disfavors the stop-NLSP scenario.
One may wonder if the large $|A_0|$ values depicted can be physical, since
typically charge and color-breaking minima occur for $|A_0|$ much larger than
$m_0$ \cite{Casas:1995pd}. However, if the Universe is in a false  
vacuum, the tunneling probability to the real minimum can be very  
small \cite{Kusenko:1996jn}. We have checked for some examples with  
$|A_0|\sim  7 m_0$ that even if CCB minimum exists, the tunneling time  
scale is longer than the age of the Universe.

\begin{figure}
  \centering
  \includegraphics[width=.5\textwidth]{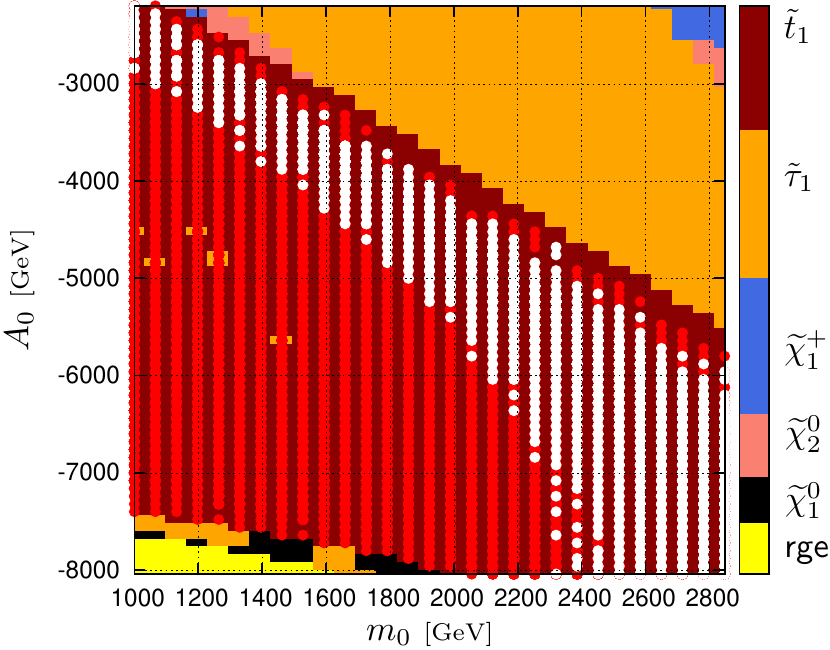}
  \caption{RD $\leq$ WMAP. Different projection.}
  \label{fig:stopOKsm0LTwmap}
\end{figure}

\begin{figure}
  \centering
  \includegraphics[width=.5\textwidth]{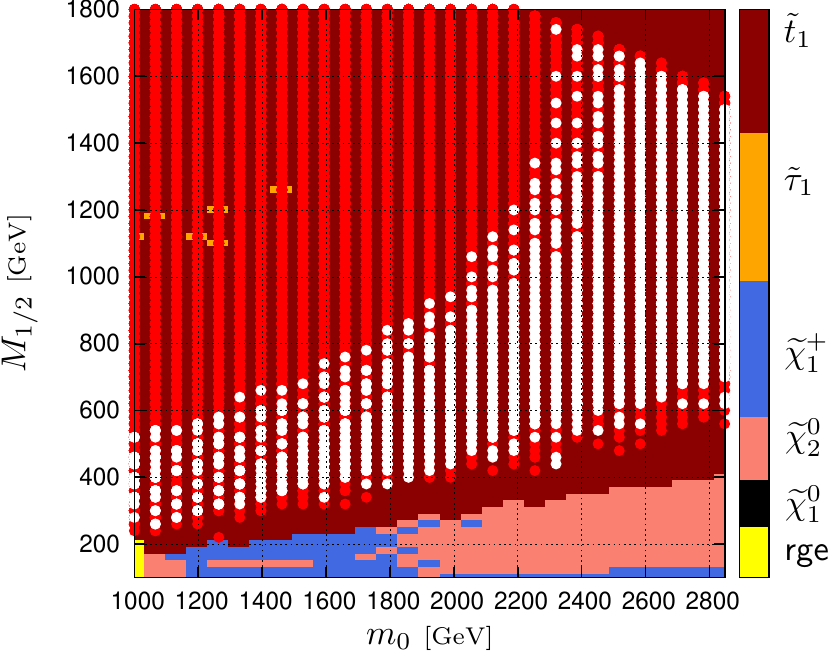}
  \caption{RD $\leq$ WMAP. Different projection.}
  \label{fig:stopOKsm0m12LTwmap}
\end{figure}

Figure~\ref{fig:stopOKsm0LTwmap} shows the equivalent of
Fig.~\ref{fig:stopOKsLTwmap}, but now the points are projected to
$(m_0,A_0)$-plane.
Figure~\ref{fig:stopOKsm0m12LTwmap} shows the projection to the
$(m_0,\mhalf)$-plane.
In Video~\ref{vid:maxmix}, the $\tb=10$ slice of
Fig.~\ref{fig:stopOKsm0m12LTwmap} is plotted.  The points of Higgs
maximal mixing (dark green) are distinguished from the other points
that obey the relic density upper-limit constraint (red dots) and the
points fulfilling also the other constraints (white dots). The green
points lie within 5\% range from the optimal mixing value of
\refeq{eq:1}. (See the animation of the evolution of the trilinear
coupling from \cite{vid}.) The maximal mixing coincides well with
the good points, suggesting that the Higgs mass limit is, in fact,
quite a severe constraint.
The large area of sufficiently low relic density points (red) is
generated, when the RGE conditions push the \mhalf
parameter larger as the absolute value of the trilinear-parameter
increases. This can be clearly seen in the video.
\begin{video}
  \centering
  \includegraphics{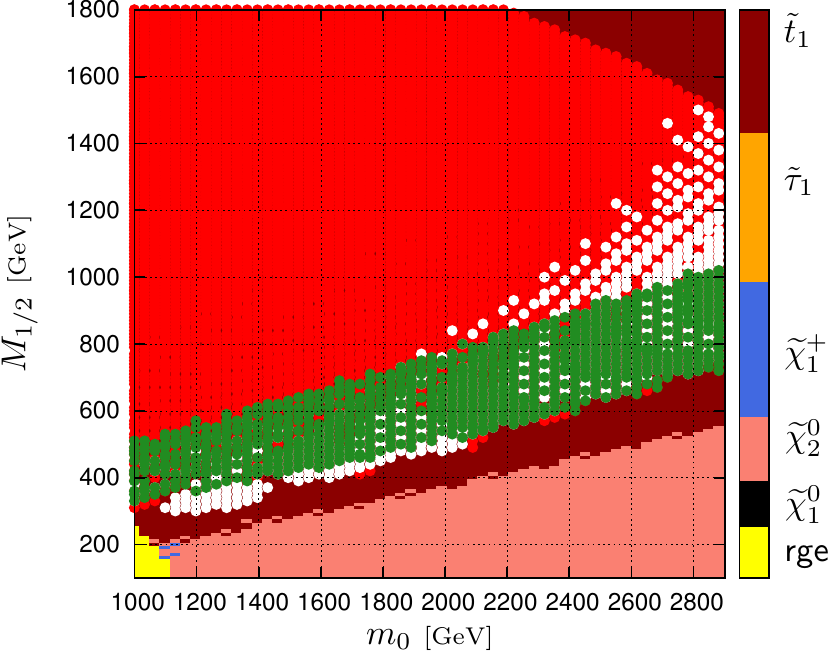}
  \setfloatlink{http://www.youtube.com/watch?v=vyizA4262Qg}
  \caption{\label{vid:maxmix}Points of optimal mixing (green) for stop
    NLSP points, $\tb =10$, RD $\leq$ WMAP. The scan is somewhat
    denser  in this plot than in Fig.~\ref{fig:stopOKsm0m12LTwmap}.}
\end{video}

Figures \ref{fig:stopOKsLTwmap}--\ref{fig:stopOKsm0m12LTwmap} are
different projections of the same parameter volume. It is interesting
to see how the viewpoint exaggerates or understates certain aspects of
the parameter space. For example, the area with no good spectrum
(denoted by rge) seems to be quite large in
Fig.~\ref{fig:stopOKsLTwmap}, whereas in
Fig.~\ref{fig:stopOKsm0m12LTwmap} it is hardly visible.
An important message to be learned from this is that 2-dimensional
extracts from a multidimensional parameter space may give a false
feeling of the parameter space being very restricted.

\subsection{Masses}
\label{sec:masses}
Searches for squarks and gluinos at the LHC disfavor a low mass region
\cite{Khachatryan:2011tk,Collaboration:2011qk}.
If only stop is light, the anticipated exclusion limit would be lower
than in the case of degenerate squarks and gluino due to smaller
production cross section, and partly also due to $t\bar{t}$
background. Also, proximity of the stop NLSP to the LSP might result
in the supersymmetric events failing the missing energy cuts.
A CMS search for jets+MET resulted in $N_{max}=13$ events at 95\%
confidence level for an integrated luminosity $\mathcal L = 35 \text{
  pb}^{-1}$. The upper bound is related to the total SUSY production
cross section by $N_\text{max}= \epsilon\mathcal L
\sigma_\text{max}$. If we assume the total efficiency $\epsilon$ to be
25\% ({\it e.g.}~see the discussion on efficiency in
\cite{Khachatryan:2011tk}), we can estimate the upper bound on the
cross section to be $\sigma_\text{max} = 1.5$~pb.
Estimating that about one picobarn production cross section is needed
to exclude a certain sparticle, we get for the lonely stop NLSP an
anticipated lower mass limit of 300 GeV from $\tilde t \tilde t^*$
production using Prospino
\cite{Beenakker:1996ed,*Beenakker:1997ut,*Beenakker:1996ch}.  However,
stop NLSP may not be easy to detect at the LHC, as will be discussed
in the next section.

\begin{figure}
  \centering
  \includegraphics[width=.5\textwidth]{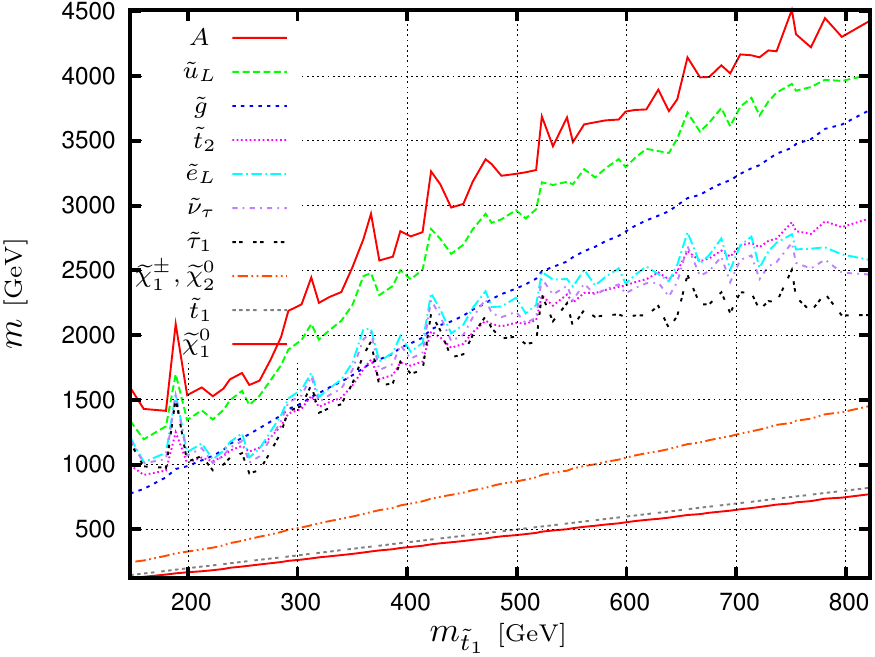}
  \caption{Mass spectrum of a set of good stop NLSP points as a
    function of the stop mass. In the key, particles are listed in
    descending mass order found at $\mst=800$ GeV (RD=WMAP).}
  \label{fig:spect}
\end{figure}

Interestingly, a small mass difference between stop and the LSP, and
also a wide gap between the lighter stop and the other squarks are
exactly what we find in the case of stop NLSP.
Large splitting in the stop sector pushes lighter stop down to be a
lonely SUSY scalar. The other scalars, except the light Higgs, are
much heavier, with masses typically above 1 TeV.
This is because the stop NLSP prefers large $m_0$ as compared to
\mhalf (about twice the \mhalf value in the following).
In Fig.~\ref{fig:spect}, a typical mass spectrum is plotted as a
function of \mst. Only one stop and \Neu 1 are light, with masses
close to each other. Each point in the plot is selected from the white
points of the scan in Sec.~\ref{sec:stop-nlsp-regions}.
The stop mass follows quite faithfully the \mhalf parameter, but the
other parameters do not increase monotonically with the stop mass for
the selected points. Therefore the neighboring points are not
necessarily next to each other in the 4-dimensional parameter space.
They, however, obey all the constraints except $a_\mu$, and were also
required to explain the amount of dark matter (relic density is within
the limits of \refeq{wmaplimits}, which we denote as RD=WMAP).
\begin{figure}
  \centering
  \includegraphics[width=.5\textwidth]{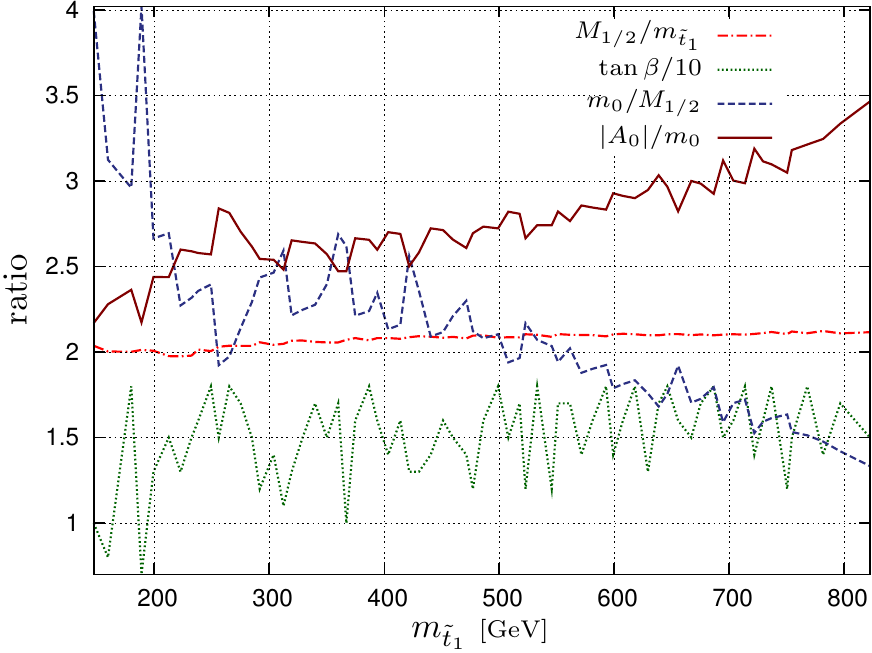}
  \caption{Values of the parameters for Fig.~\ref{fig:spect}}
  \label{fig:params}
\end{figure}
In Fig.~\ref{fig:params}, we show the ratios of several parameters for
the points in Fig.~\ref{fig:spect} (and consequently, for
Figs.~\ref{fig:st1BRs}--\ref{fig:cstb}).  In particular, it turns out
that for these points, $|A_0| < 3.5 m_0$ always. The favored value of
$\tan\beta \sim 15$. It is also clear that for heavier stops, $m_0$
and $M_{1/2}$ come closer to each other, while $|A_0|$ value increases
compared to $m_0$.

\begin{figure}
  \centering
  \includegraphics[width=.5\textwidth]{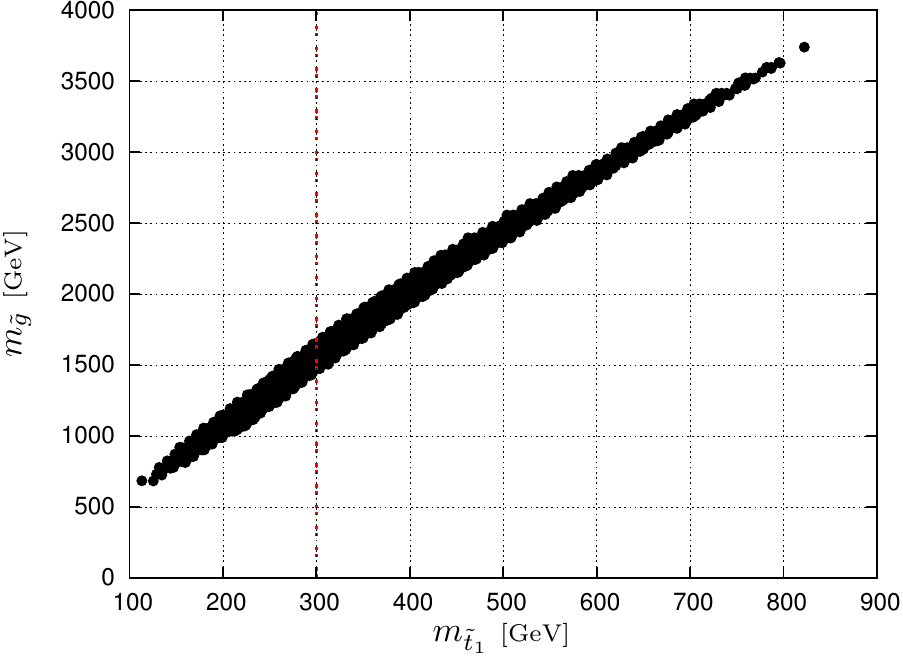}
  \caption{Allowed points in (\mst, \mgl) plane, with stop NLSP, RD
    $\leq$ WMAP.  Vertical line reminds about the assumed LHC stop
    mass limit.}
  \label{fig:stVSmgOKStNLSPLEQ}
\end{figure}

Figure~\ref{fig:stVSmgOKStNLSPLEQ} shows the allowed stop NLSP points
in the
$(m_{\tilde t_1},m_{\tilde g})$ plane for RD $\leq$ WMAP. These points
are the white points of the scan of Sec.~\ref{sec:stop-nlsp-regions}.
There is a strong correlation with stop and gluino masses. If the stop
NLSP mass is required to exceed 300 GeV, it would imply that the
gluino mass is rather heavy. This is easily understood, since the LSP
is a bino in a large part of the CMSSM parameter space. The relic
density constraint requires the stop mass therefore to be rather close
to the \Neu 1 mass, so that co-annihilations are able to dilute the
excess LSP density. The gaugino mass relation $M_1:M_2:M_3 \simeq
1:1.9:6.2$ at the EW-scale \cite{Huitu:2010me} then predicts the
gluino mass to be about 5--6 times the stop mass ($\mgl\approx 6.2
\times M_1$, and $\mst=m_{LSP}+10\%=1.10 \times M_1$, hence
$\mgl/\mst 
\approx 5.5$).

\begin{figure}
  \centering
  \includegraphics[width=.5\textwidth]{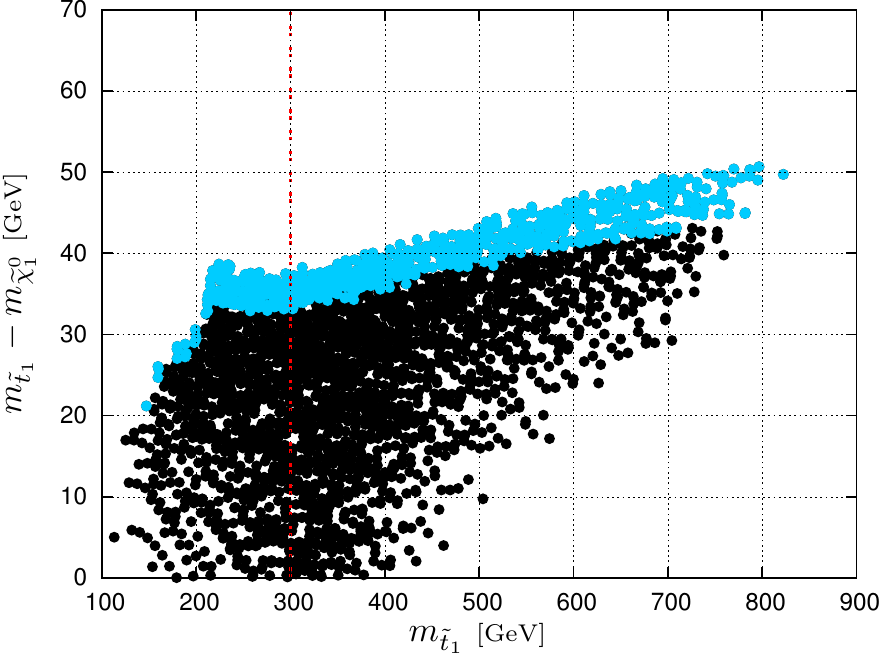}
  \caption{Mass difference of the two lightest SUSY particles
    wrt. \mst, for black dots RD $<$ WMAP and for light blue dots
    RD=WMAP.  Vertical line reminds about the assumed LHC stop mass
    limit.}
  \label{fig:mst-lsp_stNLSPoksLEQ}
\end{figure}

In Fig.~\ref{fig:mst-lsp_stNLSPoksLEQ}, the mass difference of
NLSP-stop and \Neu 1 is plotted with respect to the stop mass for the
same allowed stop NLSP points.  The mass difference seems to be
typically below approximately 50 GeV. This is a consequence of the relic density
constraint, which requires effective co-annihilations for the bino
LSP. (The exact WMAP preferred region saturates the upper edge of the
mass range, see Fig.~\ref{fig:mst-lsp_stNLSPoksLEQ}.) Another implication
is that the stop decay channels are limited to the ones explained in
Sec.~\ref{sec:stop-as-an}.

\section{Stop NLSP at the LHC}
\label{sec:at-lhc}

In the stop NLSP case, it is quite possible that most of the
supersymmetric partners are relatively heavy as discussed in the
previous section.  Discovering stop has been studied in detail in
several works, {\it e.g.}, in
\cite{Bhattacharyya:2008tw,Bornhauser:2010mw,Johansen:2010ac,Hiller:2009ii,Kraml:2005kb,Das:2001kd,
  Olive:1994qq,Demina:1999ty,Abel:2000vs,Bhattacharyya:2011ew}.  Here,
we will shortly discuss the cross sections and decay modes of stop
NLSP at the LHC, when the constraints discussed in previous sections
are fulfilled.

Typically, the $\tilde t_1 \tilde t_1^*$ cross section is very large
compared to the other production mechanisms via cascade decays of
other sparticles.  For example, in Table~\ref{tab:cs}, the
next-to-leading order LHC cross sections for several squark and
gaugino production channels are calculated for one acceptable
parameter point corresponding to $\mst=304$ GeV of
Fig.~\ref{fig:spect}.
\begin{table}
  \centering
  \begin{tabular}[c]{|c|c|c|}
    \hline 
    $\sigma(pp\to \tilde x\tilde y)$ & 7 TeV & 14 TeV\\
    \hline
$\tilde t_1 \tilde t_1^*$ 
& 1130 & 9240 \\
$ \tilde{g} \tilde q$ 
& $ 0.958 \times 10^{-1}$ & 30.2\\
$ \Neu 2 \Chap 1 $
& 3.36 & 21.5 \\
$\tilde{g}\tilde{g}$
& $0.634 \times 10^{-1}$ & 15.7 \\
$\Chap 1 \Cham 1$
& 2.05 & 14.9 \\
$\tilde q \tilde q$
& $0.184 \times 10^{-1}$ & 9.48  \\
$\Neu 2 \tilde g$
& $0.136\times 10^{-1}$ & 0.679 \\
$\Chap 1 \tilde q$
& $0.969 \times 10^{-2}$ & 0.639 \\
$\Neu 2 \tilde q$
& $ 0.635\times 10^{-2}$ & $ 0.452$ \\
\hline
  \end{tabular}
  \caption{Cross sections [fb] at the LHC for  $\mhalf = 620$ GeV, $m_0 =
    1528$ GeV, $A_0 = -3880$ GeV, $\tb = 14$, and sign$(\mu)=+1$, 
    corresponding to $\mst = 304$ GeV of
    Fig.~\ref{fig:spect} (Prospino2, NLO).}
  \label{tab:cs}
\end{table}

\begin{figure}
  \centering
  \includegraphics[width=.5\textwidth]{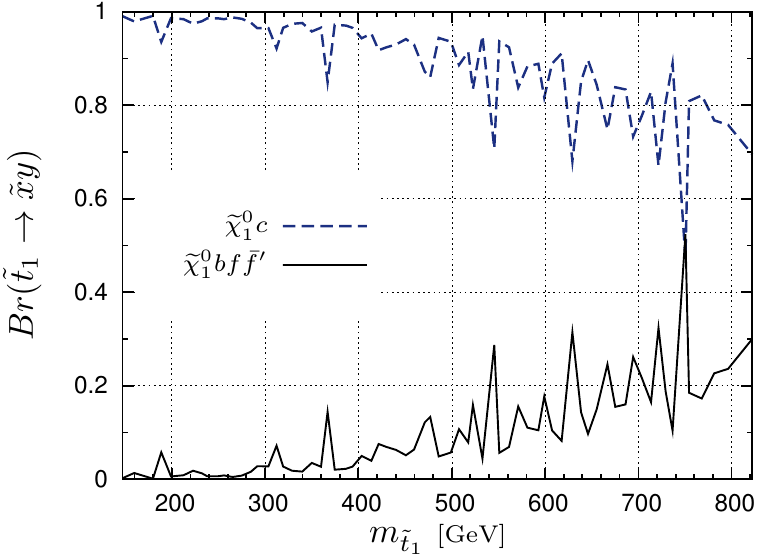}
  \caption{Branching ratios of lightest stop for the collection of
    allowed points, RD $=$ WMAP.}
  \label{fig:st1BRs}
\end{figure}

\begin{figure}
  \centering
  \includegraphics[width=.5\textwidth]{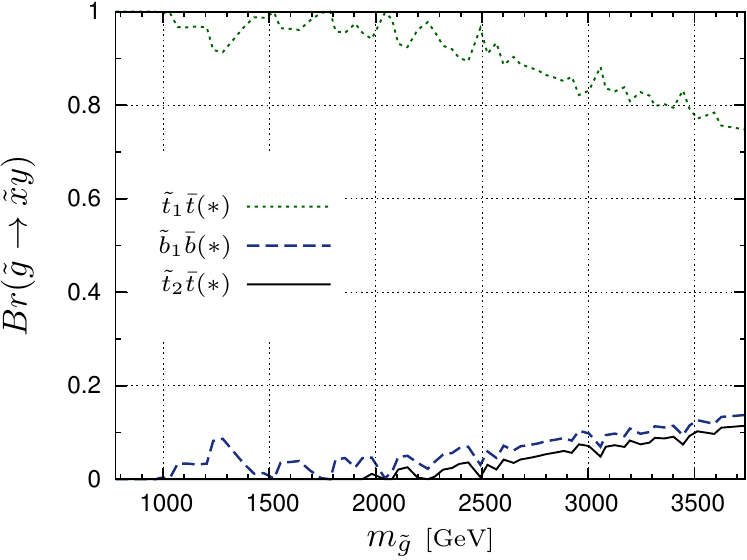}
  \caption{Branching ratios of gluino for the collection of allowed
    points, RD $=$ WMAP. $(*)$ indicates that the charge conjugate
    state is also included.}
  \label{fig:gluiBRs}
\end{figure}

In Fig.~\ref{fig:st1BRs}, we have plotted the dominant decay modes of
the lightest stop as a function of its mass for the points in
Fig.~\ref{fig:spect}.  The dominant decay mode is through $\Neu 1 c$
channel, but the importance of $bf\bar f'\Neu 1$ channel increases
with increasing stop mass and the mass difference to the LSP
(Fig.~\ref{fig:mst-lsp_stNLSPoksLEQ}).
In Fig.~\ref{fig:gluiBRs}, the branching ratios of gluino are plotted
as a function of its mass for the same points.  Gluino decays
dominantly to a stop-top pair.  The charge conjugated mode is also
included in the plotted value.
For the evaluation of branching ratios and the decays of the
supersymmetric particles, we have used SUSY-HIT (v.1.3 with SDECAY
v1.3b/HDECAY v3.4)~\cite{Djouadi:2006bz}.
It should be noted, though, that the approximate result for the $\stp
\to \Neu 1 c$ decay used in
~\cite{Djouadi:2006bz} 
has to be taken with care in case the minimal flavor violation scale
is not large and has to be reanalyzed more carefully
\cite{Muhlleitner:2011ww}.
The cross sections have been
calculated to next-to-leading order (NLO) using Prospino2 (v.''\verb
on_the_web_11_3_10'')
\cite{Beenakker:1996ed,*Beenakker:1997ut,*Beenakker:1996ch}.

\begin{figure}
  \centering
  \includegraphics[width=.5\textwidth]{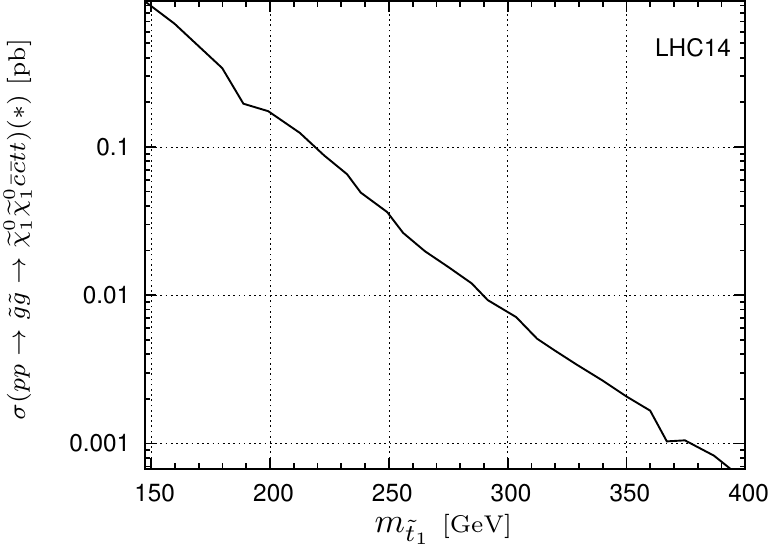}
  \caption{Cross section of the process $pp\to \tilde g \tilde
    g\to\Neu 1 \Neu 1 \bar c \bar c tt (*)$ [pb],
    for collection of allowed points, RD $=$ WMAP. The final state
    includes both combinations of like-sign tops.}
  \label{fig:csgg}
\end{figure}

In Fig.~\ref{fig:csgg}, the LHC 14 TeV cross section for the like-sign
top process $pp\to \tilde g \tilde g\to\Neu 1 \Neu 1 \bar c \bar c
tt/\Neu 1 \Neu 1 \bar t \bar t cc$ [pb] has been plotted as a function
of stop mass up to 400 GeV, where the cross section is
already hopelessly small.  The points correspond to the previously
used stop NLSP set, which passes the constraints (points in
Figs.~\ref{fig:spect}, \ref{fig:st1BRs}, \ref{fig:gluiBRs}). The
plotted cross section contains contributions from both charge
conjugation final states.  For $m_{\tilde t_1}=300$ GeV, the cross
section is around 10 fb but decreases fast with increasing stop mass,
which is caused by the declining gluino production cross section.  The
background of like-sign decays can be removed \cite{Kraml:2005kb},
which makes the process interesting for probing light stops at the
LHC.

\begin{figure}
  \centering
  \includegraphics[width=.5\textwidth]{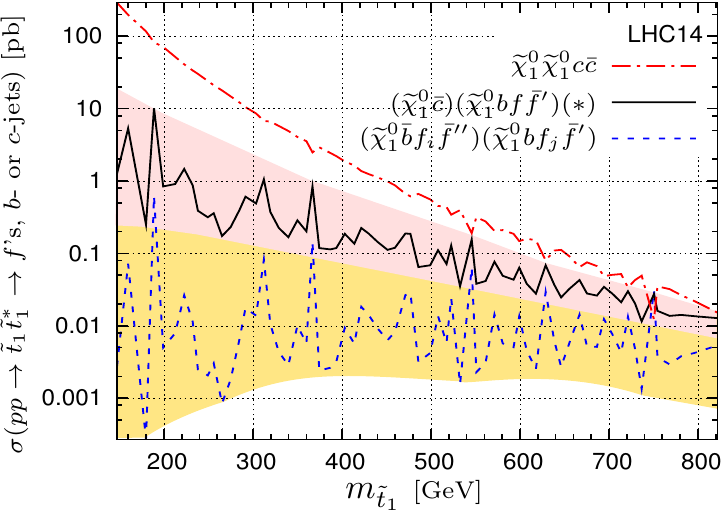}
  \caption{Cross section of the process $pp\to \tilde t_1 \tilde
    t_1^*\to \Neu 1 \Neu 1 c \bar c$ [pb] (upper), the process $pp\to
    \tilde t_1 \tilde t_1^*\to (\Neu 1 \bar c) (\Neu 1 b f \bar f')$
    [pb] (in the middle, including the charge conjugated final state),
    and the process $pp\to \tilde t_1 \tilde t_1^*\to (\Neu 1 \bar b
    f_i\bar f'') (\Neu 1 b f_j \bar f')$ [pb] (the lowest), for
    collection of allowed points, RD $=$ WMAP. }
  \label{fig:cstb}
\end{figure}

The stop pair production cross section is clearly the largest
production channel. However, for a light stop, the loop decay
dominates, and the final state of almost back-to-back neutralinos and
two soft $c$-jets is not experimentally promising.  For heavier stops,
the four-body branching ratio competes with the loop decay one.  In
Fig.~\ref{fig:cstb}, the LHC 14 TeV cross section for the process
$pp\to \tilde t_1 \tilde t_1^*\to (\Neu 1 b f \bar f')( \Neu 1 \bar
c)+$charge conjugated final state [pb] has been plotted for the same
stop NLSP set, which passes the constraints (the middle curve).  The
cross section remains reasonably large, $\cal{O}$(10 fb), even for
$m_{\tilde t}\sim 800$ GeV.  When the fermions $f f^\prime$ are $\ell\nu$
from $W$, the signal would be a $b$-jet and charged lepton in one
hemisphere and missing energy from neutralinos, neutrinos, and a soft
$c$-jet in the other hemisphere.  A detailed signal analysis is not
the purpose of this work, but this signature may be possible.  The
cross section where both stops decay to four particles is for all
studied stop masses around $\cal{O}$(10 fb), although the variation
between neighboring points may be large.

\section{Summary and discussion}
\label{summary}
We have studied the interesting possibility of stop being the
next-to-lightest supersymmetric particle within the CMSSM
scenario. 
Large mixing in the stop mass matrix can result in a (relatively)
light stop squark, and in some cases, it can be the NLSP. Typically
this prefers a large nonzero value for the trilinear mass parameter
$A_t$.  In addition, a large $\mu$-parameter naturally splits the stop
pair, and may also result in a light \mst, unless there is
cancellation between $\mu$ and $A_t$.
Large parameter spaces with the lighter stop as an NLSP can be found,
which simultaneously fulfill various experimental constraints. Within
these regions, the neutralino LSP may explain the dark matter problem.
The \Btn, \Bsg and \Bmumu constraints agree with the experimental
limits, but we have not found points where the possibly large observed
discrepancy of \gmu from the SM expected value could be explained.
The CP-even Higgs boson lower mass limit is a severe constraint, which
many times rules otherwise good parameter spaces out.

The LHC is currently pushing squark and gluino lower mass limits to
higher values. The nonobservation of sparticles may fit to the
scenario, where the NLSP is a rather heavy squark. In CMSSM, this would
in most cases be the lighter stop squark.
We have argued that observing the top signal resulting from this
scenario is quite challenging.
The like-sign top channel, that relies on the Majorana character of
the gluinos, requires reasonably large production cross section of
gluinos at the LHC. In the stop NLSP scenario with, let us say, heavier
than 350 GeV stop, the gluino is very heavy, about five times that, by
the WMAP constraint and gaugino mass ratios. Therefore, the gluino
production will be scant, resulting in the like-sign top signal to lie
in a subfemtobarn range.  Because of the relative lightness of stop with
respect to the other SUSY scalars, their direct pair production
dominates.  In the case of heavy stop NLSP, the production and decay
chain $pp\to\tilde t_1\tilde t_1^*\to b\ell \nu +E\!\!\!/_T$ may be
possible (see also \cite{Bhattacharyya:2011ew}).  Alternatively to the
stop production alone, signals where other particles are used in order
to discover stops can be utilized, as studied, {\it e.g.}, in
\cite{Bornhauser:2010mw} for light stops.

\section{Acknowledgments}
We thank Ritva Kinnunen for discussions, and Debajyoti Choudhury for
participating in the early stage of the investigation.  We thank
Nordita program {\it TeV Scale Physics and Dark Matter} for
hospitality while this work was being initiated.  KH and LL
acknowledge support from the Academy of Finland (Project No.  137960).
The work of JL~is supported by the Foundation for Fundamental Research
of Matter (FOM).  LL thanks Magnus Ehrnrooth Foundation for financial
support.

\bibliographystyle{apsrev4-1}
\bibliography{stopnlsp}
\end{document}